\documentclass[aps,prl,twocolumn,floats,longbibliography,showkeys,nofootinbib]{revtex4-1}
\usepackage[dvips]{graphicx}

\usepackage{amsmath}
\usepackage{braket}
\usepackage{amsfonts}
\usepackage[colorlinks=true]{hyperref}
\usepackage{hyperref}
\usepackage{xcolor}

\usepackage{booktabs}

\usepackage{aas_macros}

\usepackage{caption}
\usepackage{graphicx}
\usepackage{subcaption}

\usepackage{bm}
\usepackage{slashed}

\def\aap{\textit{Astronomy and Astrophysics}}

\def\be{\begin{equation}}
\def\ee{\end{equation}}
\def\bea{\begin{eqnarray}}
\def\eea{\end{eqnarray}}

\newcommand\D{\mathcal D}

\def\L{\mathcal{L}}

\begin{document}

\title{Spacetime might not be doomed after all}

\author{Olivier Minazzoli}
\email[]{ominazzoli@gmail.com}
\affiliation{Artemis, Universit\'e C\^ote d'Azur, CNRS, Observatoire C\^ote d'Azur, BP4229, 06304, Nice Cedex 4, France,\\Bureau des Affaires Spatiales, 2 rue du Gabian, 98000  Monaco}

\begin{abstract}
Popular wisdom amongst theoretical physicists says that the continuum structure of spacetime is probably not elementary, but rather emergent. While many arguments to support that view arise from speculative ideas, the argument can also be made by only invoking standard physics. In this manuscript, I shall argue that a novel general theory of relativity might change the deal, while it corresponds to a somewhat minimal extension of the core theory of physics. 
\\ \center \today
\end{abstract}
\maketitle

During the last few decades, there has been a growing rumor amongst theoretical physicists and philosophers of science that he usual continuum structure of spacetime cannot be fundamental after all, and that it probably actually emerges from more elementary constituents. Some even say, with striking confidence, that spacetime is \textit{doomed} \cite{arkani-hamed}. Although many arguments against the elementary nature of spacetime seem to arise from speculative areas of theoretical physics \cite{gross:2008pu,ashtekar:2004cq}, it is possible to argue in that direction directly from the path integral formulation of the core theory of physics---that is, general relativity plus the standard model of matter \cite{wilczek:2016bk}.
Indeed, the path integral formulation of the core theory of physics reads
\be
Z_{\textrm{C}} = \int \D g \prod_i \D f_i \exp \left[\frac{i}{\hbar c} \int d^4_g x \left(\frac{R(g)}{2 \kappa} + \L_m(f,g) \right)\right], \label{eq:corePI}
\ee
where $f_i$ are the matter fields of the standard model of particles---such as fermions and gauge bosons, and the Higgs---whereas $g$ stands for the metric, and $d^4_g x = \sqrt{-g} d^4x$ is the elementary spacetime volume. $R$ is the Ricci scalar constructed upon the metric, and $\L_m(f,g)$ is the standard model of particles that depends on both the fields themselves and the spacetime metric according to the \textit{comma-goes-to-semicolon rule} \cite{MTW}. For the purpose of the argument, I do not restrict the path integral to energies below a cutoff scale.
There are three universal constants in this formulation: the quantum constant $\hbar$ (Planck's), the causal structure constant $c$ \footnote{Which turns out to be equal to the speed of light in vacuum on scales for which spacetime can be approximated to be flat.} and the constant of gravity $G = c^4 \kappa / (8\pi)$ (Newton's). From these constants, one can construct an energy scale, a mass scale, a time scale and a length scale, known as the Planck energy ($E_P$), mass ($m_P$), time ($t_P$) and length ($l_P$) respectively:
\be
E_P = \sqrt{\frac{\hbar c^5}{G}}, m_P = \frac{E_P}{c^2}, t_P = \sqrt{\frac{\hbar G}{c^5}}, l_P = c t_P.
\ee
One can directly see the fundamental role of the Planck length for pure quantum general relativity from its path integral formulation 
\be
Z_{\textrm{GR}} = \int \D g \exp \left[\frac{i}{2 l_P^2} \int d^4_g x R(g)\right], \label{eq:GRPI}
\ee
for which the Planck length squared $l_P^2$ plays the role of a quantum of area, the same way Planck's constant $\hbar$ plays the role of a quantum of action in standard quantum field theory and quantum mechanics.

From various theoretical arguments \cite{mead:1964pr,isham:1994ln,garay:1995ij,bekenstein:2008sc,dyson2013ij}, the Planck length and time are thought to be the elementary units of spacetime. And it turns out that all the tough questions associated to quantum general relativity are related one way or another to these elementary units \cite{hawking:1978np,isham:1994ln,loll:2008cq}, as the Planck time and space often signify the scales at which something fundamentally transformative seems to happen to the structure of spacetime \cite{isham:1994ln,ashtekar:2004cq,loll:2008cq}.

Is there any way to define a quantum field theory of gravity that does not imply the existence of elementary scales for space and time?
I will argue that yes, there is. Furthermore, I will explain that this alternative does not require any exotic and highly speculative ideas such as extra dimensions, compactification, branes, etc. Let us define the path integral formulation of an alternative theory to the core theory that reads as follows
\be
Z_{\textrm{ER}} = \int \D g  \prod_i \D f_i \exp \left[-\frac{i}{2 \epsilon^2} \int d^4_g x \frac{\L^2_m(f,g)}{R(g)} \right], \label{eq:ERPI}
\ee
where $\L_m(f,g)$ might still be the standard model of particles, but most likely is an unknown completion of it. $\epsilon$ is a constant to be determined, with the units of an energy. This weird looking quantum field theory remains to be studied, but one can check the broad consequences of its semi-classical limit, for which the background is assumed to be classical. Indeed, at the background level, the following actions are equivalent
\bea
-\frac{1}{2\epsilon^2} \int d^4_g x \frac{\L^2_m(f,g)}{R(g)} &&\equiv \label{eq:equiv}\\ 
&&\frac{1}{\epsilon^2} \int d^4_g x \frac{1}{\kappa}\left(\frac{R(g)}{2 \kappa} + \L_m(f,g) \right), \nonumber
\eea
provided that $\L_m \neq \emptyset$, and where $\kappa$ is now a scalar-field instead of being a constant as in Eq. (\ref{eq:corePI}). Obviously, the value of $\epsilon$ does not impact the classical limit of the theory. The equivalence between the original \textit{$f(R,\L_m)$} theory in the left-hand-side of Eq. (\ref{eq:equiv}) and the \textit{Einstein-dilaton} theory in the right-hand-side, stems from a very well known fact: non-linear algebraic functions of the Ricci scalar in the action are equivalent to having an additional scalar degree-of-freedom with gravitational strength \cite{capozziello:2015sc}. As a consequence, it indicates that the weird looking theory in Eq. (\ref{eq:corePI}) should be immune to the Ostrodrogradskian instability and the Cauchy problem despite not being of second order---just as $f(R)$ theories \cite{woodard:2007ln,teyssandier:1983jm,*jakubiec:1988pr}. Eq. (\ref{eq:equiv}) also shows that, at the classical level, Newton's constant $G := \kappa c^4 / (8\pi)$ is actually not a constant but a field. Otherwise, whereas $\kappa = - R / T$ in general relativity, one can check that $\kappa = - R / \L_m^o$ from the right-hand-side of Eq. (\ref{eq:equiv}), where $\L_m^o$ is the on-shell value of $\L_m$. As usual in $f(R)$ theories, the trace of the metric field equation yields to an equation on the gravitational scalar degree-of-freedom, which here reads
\be
3 \kappa^{2} \square \kappa^{-2}=\kappa\left(T-\mathcal{L}_{m}^o\right).\label{eq:SF}
\ee
Hence, whenever $\L_m^o = T$, the scalar degree-of-freedom is not sourced. The consequence of this \textit{intrinsic decoupling} \cite{minazzoli:2013pr,minazzoli:2014pr,minazzoli:2014pl} is that the theory behaves as general relativity---that is $\kappa \sim$ constant---pretty much whenever $\L_m^o \sim T$, which actually occurs in many expected situations of the observable universe. It follows that, at the classical level, this theory predicts that the post-Newtonian parameters $\gamma$ and $\beta$ are both equal to one \cite{minazzoli:2013pr,minazzoli:2016pr,*bernus:2022pr}---as in general relativity---that the value of Newton's constant $G$ freezes at least at the beginning of the matter era \cite{minazzoli:2014pr,*minazzoli:2014pl,minazzoli:2021cq}, that neutron stars are not much different from the ones of general relativity \cite{arruga:2021pr,arruga:2021ep}, nor seem to be black-holes \cite{minazzoli:2021ej}, nor even gravitational-waves \cite{hirschmann:2018pr,khalil:2018pr}. The trace Eq. (\ref{eq:SF}) of the metric field equation imposes a differential condition on the ratio between $\L_m^o$ and $R$, the same way the trace of Einstein's equation imposes an algebraic condition on the ratio between $T$ and $R$. These ratio simply give the amplitude with which matter curves spacetime at the classical level, either in entangled relativity or in general relativity. Indeed, the metric field equation from both sides of Eq. (\ref{eq:equiv}) reads \cite{ludwig:2015pl,*harko:2013pr}
\be
R_{\mu \nu} - \frac{1}{2} g_{\mu \nu} R = \kappa T_{\mu \nu} + \kappa^2 \left(\nabla_\mu \nabla_\nu - g_{\mu \nu} \Box \right) \kappa^{-2}.
\ee

The \textit{intrinsic decoupling} mentioned above implies that either the background value of $\kappa$ is a constant most of the time in the observable universe, or it varies even less than the spacetime metric. Hence, for any quantum phenomenon for which gravity can be neglected, the variation of $\kappa$ can also be neglected, such that it can be factorized outside the integral on the right-hand-side of Eq. (\ref{eq:equiv}), and one therefore gets
\be
Z_{\textrm{ER}} \sim \int \prod_i \D f_i \exp \left[\frac{i}{\kappa \epsilon^2} \int d^4 x  \L_m(f) \right], \label{eq:ERPIapp}
\ee
at the scales for which gravity can be neglected. This turns out to be nothing but standard quantum field theory ``on flat spacetime'',\footnote{Spacetime is never flat in the core theory of physics, but its variation can usually be neglected on small enough time and space scales, and for perturbations that are well below the Planck energy.} with $\hbar c = \kappa \epsilon^2 $. From the constant $\kappa$ limit of the theory, one therefore deduces that the only parameter of the theory in Eq. (\ref{eq:ERPI}) is the Planck energy $\epsilon = \sqrt{c\hbar/\kappa}$. More importantly, it means that standard quantum field theory ``on flat spacetime'' is a specific limit of the quantum field theory defined in Eq. (\ref{eq:ERPI}). Moreover, neither Planck's constant $\hbar$, nor Newton's constant $G$, are actually constant---as both can vary and are proportional to each other ($\hbar \propto G$) at the semi-classical limit for which the background value of $\kappa$ does not vary much. 
It also implies that the weak gravity limit $G\rightarrow 0$ corresponds to the classical limit $\hbar \rightarrow 0$, literally.

But only two universal constants appear in Eq. (\ref{eq:ERPI}): the Planck energy $\epsilon$, at which one expects the theory to be fully quantum, and the causal structure constant $c$. Hence, there is not enough universal constants in order to construct an elementary length scale or time scale. Therefore, there is no reason to assume that something special happens to spacetime at the Planck energy scale, as it was the case with the core theory of physics.

As a consequence, if quantum gravity is defined by Eq. (\ref{eq:ERPI}), it might mean that the continuum picture of spacetime is valid at any scale. However, it is important to stress that the behavior of the theory will likely highly depends on the definition of $\L_m$. And one can expect that the non-linear form of Eq. (\ref{eq:ERPI}) will impose strong theoretical constraints on what $\L_m$ can be for the theory to be consistent at the full quantum level. All that we know for now, is that it must reduce to the standard model of particles in the limit of Eq. (\ref{eq:ERPIapp})---and when the value of $G$ is precisely its measured value on Earth.

But the good thing with this theory is that it might be tested in a very specific way. Indeed, even if the intrinsic decoupling discussed above implies that the background value of $\kappa$ is pretty much a constant everywhere in the observable universe, it should not always be a constant. Indeed, for instance, this decoupling does not work for (unbounded \footnote{When a magnetic (or electric) field is bounded, such as in a nucleus for instance, then it contributes to the total trace of the bound object due to the constraint that the internal stresses all vanish \cite{nitti2022x}.}) magnetic fields, since $\L_m^o (\propto B^2) \neq T =0$ for a magnetic field. That means that the energy of (unbounded) magnetic fields can source the equation of the gravitational scalar-field $\kappa$ Eq. (\ref{eq:SF}), although the amplitude of its (classical) perturbation will be as weak as the metric (classical) perturbation that is created by magnetic fields. Hence, one can expect $\kappa$---and therefore $\hbar$---to slightly vary for intense (or spatially large) magnetic fields---or for other situations that are yet to be found. This means that, at least in principle, this unique aspect of the theory may be probed at the experimental or observational level. Whether or not that will be feasible in a near future is another question.

\section*{Afterword}

The theory defined in Eq. (\ref{eq:ERPI}) has been named \textit{entangled relativity} in \cite{arruga:2021pr}. Not because it appeared to be related to \textit{quantum entanglement}, \textit{a priori}, but because the theory could not even be defined without defining matter fields at the same time. Such that, unlike in general relativity, gravity and matter could not be treated separately at the fundamental level: the theories of gravity and of matter fields are \textit{entangled} in that sense. But this also means in turns, that Eq. (\ref{eq:ERPI}) forbids \textit{inertia} to be defined without matter fields. Such that, entangled relativity not only reduces to the core theory of physics in the $\kappa=$ constant limit, but also that, unlike general relativity, it satisfies the \textit{principle of relativity of inertia} \cite{einstein:1917ap,einstein:1918ap,einstein:1918an,pais:1982bk,book_mach_principle,hoefer:1995cf} that Einstein named \textit{Mach's principle} in \cite{einstein:1918an}. Indeed, Eq. (\ref{eq:ERPI}) forces spacetime to be permeated with matter fields, such that non-rotating frames simply refers to frames in which the background of matter fields is non-rotating. On the contrary, in general relativity, inertial frames could very well be defined without any reference to any form of matter fields---as one can see from Eq. (\ref{eq:GRPI}). Hence, general relativity does not enforce the \textit{principle of relativity of inertia} \cite{hoefer:1995cf}, whereas entangled relativity does.

\bibliography{ER_doom}

\end{document}